\documentclass[journal]{IEEEtran}
\usepackage{ntheorem}
\usepackage{amsmath}
\usepackage{amsfonts}
\newtheorem{theorem}{Theorem}

\newtheorem{lemma}{Lemma}

\newtheorem{assumption}{Assumption}
\newtheorem{remark}{Remark}
\usepackage{ntheorem}
\usepackage{amsmath}
\usepackage{amsfonts}
\usepackage{listings}
\usepackage{color}
\usepackage{array}
\usepackage{tabu}
\usepackage{mathrsfs,amssymb}
\usepackage{graphicx}
\graphicspath{ {image22/} }
\usepackage{float}
\usepackage{algorithm}

{}

\usepackage{algpseudocode}
\algdef{SE}[DOWHILE]{Do}{doWhile}{\algorithmicdo}[1]{\algorithmicwhile\ #1}%
\DeclareMathOperator*{\argmin}{arg\,min}

\DeclareMathOperator*{\card}{card}
\DeclareMathOperator*{\supp}{supp}

\DeclareMathOperator*{\rank}{rank}
\begin{document}

\title{An Unknown Input Multi-Observer Approach for Estimation and
Control under Adversarial Attacks}
		
%
%
%

\author{Tianci~Yang,
	~Carlos~Murguia,
	~Margreta~Kuijper, and~Dragan~Ne\v{s}i\'{c}
	\thanks{This work was supported by the Australian Research Council under the Discovery Project DP170104099.}
	\thanks{The authors are with the Department of Electrical and Electronics Engineering, the University of Melbourne, Australia.
		{\tt\small tianciy@student.unimelb.edu.au}}
}
%
%

\markboth{Journal of \LaTeX\ Class Files,~Vol.~14, No.~8, August~2015}%
{Shell \MakeLowercase{\textit{et al.}}: Bare Demo of IEEEtran.cls for IEEE Journals}
%



\maketitle

%
%
\begin{abstract}
We address the problem of state estimation, attack isolation, and control of discrete-time linear time-invariant systems under (potentially unbounded) actuator and sensor false data injection attacks. Using a bank of unknown input observers, each observer leading to an exponentially stable estimation error (in the attack-free case), we propose an observer-based estimator that provides exponential estimates of the system state in spite of actuator and sensor attacks. Exploiting sensor and actuator redundancy, the estimation scheme is guaranteed to work if a sufficiently small subset of sensors and actuators are under attack. Using the proposed estimator, we provide tools for reconstructing and isolating actuator and sensor attacks; and a control scheme capable of stabilizing the closed-loop dynamics by switching off isolated actuators. Simulation results are presented to illustrate the performance of our tools.
\end{abstract}
		

	\author{Tianci~Yang,
		~Carlos~Murguia,
		~Margreta~Kuijper, and~Dragan~Ne\v{s}i\'{c}
		\thanks{This work was supported by the Australian Research Council under the Discovery Project DP170104099.}
		\thanks{The authors are with the Department of Electrical and Electronics Engineering, the University of Melbourne, Australia.
			{\tt\small tianciy@student.unimelb.edu.au}}
	}
	
	\markboth{Journal of \LaTeX\ Class Files,~Vol.~14, No.~8, August~2015}%
	{Shell \MakeLowercase{\textit{et al.}}: Bare Demo of IEEEtran.cls for IEEE Journals}

\section{Introduction}

Networked Control Systems (NCSs) have received considerable attention in recent years due to their numerous advantages (e.g., reduced weight, volume and installation costs, and better maintainability) when compared with traditional control systems where sensors and actuators communicate through point-to-point (wired) links. Networked Control Systems are being used in many engineering applications, e.g., energy, transportation, military, health care, and manufacturing. With the growth of NCSs, new security challenges have become an important issue as wireless communication networks increasingly serve as new access points for adversaries trying to disrupt the process. Cyber-physical attacks on NCSs have caused substantial damage to a number of engineering systems. A well-known example is the StuxNet virus that targeted Siemens' supervisory control and data acquisition systems. Another example is the false data injection attacks on power systems \cite{Liu2009}. A more recent incident happend in 2014, where the computers of a German steel mill were hacked and a destruction of a blast furnace was caused. These and many other recent incidents show that tools to identify and deal with attacks on NCSs are needed.
\\[1mm]
In \cite{Fawzi2014a}\nocite{Vamvoudakis2014}\nocite{Chong2016b}\nocite{Shoukry2014}\nocite{Teixeira2012b}\nocite{Pasqualetti123}\nocite{Sahand2017}\nocite{Murguia2016}\nocite{Carlos_Justin4}\nocite{Chong2015}-\cite{Tang2019}, various security and privacy problems for linear control systems have been addressed and solved. In general, analysis (synthesis) tools are proposed to quantify (minimize) the performance degradation induced by different classes of attacks, e.g., false-data-injection, replay, zero dynamics, and denial-of-service. There are also some results addressing the nonlinear case. The problem of state estimation for nonlinear power systems under sensor attacks is solved in \cite{Hu2017} by using compressed sensing technique. In \cite{Kim2016a}, the authors address the problem of sensor attack detection and state estimation for uniformly observable continuous-time nonlinear systems. In \cite{Shoukry2015c}, Satisfiability Modulo Theory (SMT) solvers are used for state estimation for nonlinear differentially flat systems with corrupted sensors. In our previous work \cite{Yang2018a,Yang2018c}, the problem of state estimation and attack isolation for a class of nonlinear systems with \emph{positive-slope nonlinearities} is considered. Similar to the ideas given in \cite{Chong2015c}, we provided an observer-based estimation/isolation strategy, using a bank of circle-criterion observers, which provides a robust estimate of the system state in spite of sensor attacks and effectively pinpoints attacked sensors. Most of the existing work assume actuators to be healthy and only consider sensor attacks. There are only a few results dealing with attacked actuators. For instance, in \cite{Djouadi2015a}, the authors study the effect of actuator attacks on the performance of linear quadratic regulators. In \cite{Fawzi2012} and \cite{Showkatbakhsh2018}, the problem of state estimation under sensor and actuator attacks is addressed using compressed sensing ideas and SMT-based techniques, respectively. An adaptive control scheme that guarantees uniform ultimate boundedness of the closed-loop dynamics despite of sensor and actuator attacks is given in \cite{Yadegar2017}.\\[1mm]
The core of our estimation scheme is inspired by the work in \cite{Chong2015}, where the problem of state estimation for \emph{continuous-time LTI systems} is addressed. The authors propose a multi-observer estimator, using a bank of Luenberger observers, that provides a robust estimate of the system state in spite of sensor attacks. In this manuscript, using banks of Unknown Input Observers (UIOs), we address the problem of robust state estimation, attack isolation, and control for discrete-time LTI systems (with matrices $(A,B,C)$) under (potentially unbounded) actuator and sensor attacks. Unknown input observers are dynamical systems capable of estimating the state of the plant \textit{without} using any input signals. If such an observer exists for the matrices $(A,B,\tilde{C}_{i})$, where $\tilde{C}_{i}$ denotes a submatrix of $C$ with fewer rows and the same number of columns, then, using a bank of observers, we can perform state estimation and attack isolation when a sufficiently small subset of sensors is attacked (even if all inputs are under attack). The main idea behind our multi-observer estimator is the following. Each UIO in the bank is constructed using a triple $(A,B,\tilde{C}_{i})$, i.e., the $i$-th observer is driven by the output signals associated with $\tilde{C}_{i}$ only. If the outputs corresponding to $\tilde{C}_{i}$ are attack-free, this UIO produces an exponentially stable estimation error. For every pair of UIOs in the bank, we compute the largest difference between their estimates. Then, we select the pair leading to the smallest difference and prove that these observers reconstruct the state of the system exponentially. If a UIO does not exist for $(A,B,\tilde{C}_{i})$, but it does for $(A,\tilde{B}_{i},\tilde{C}_{i})$, where $\tilde{B}_{i}$ is a submatrix of $B$ with fewer columns and the same number of rows, i.e., the $i$-th observer does \textit{not} use the input signals associated with $\tilde{B}_{i}$, but it does use the remaining input signals and the output signals corresponding to $\tilde{C}_{i}$, then using a bank of these UIOs, we can use similar ideas to perform state estimation and attack isolation at the price of only being able to isolate when a sufficiently small subset of actuators and sensors are under attack. If the inputs corresponding to $\tilde{B}_{i}$ include all the attacked ones and the outputs corresponding to $\tilde{C}_{i}$ are attack-free, this UIO produces exponentially stable estimation error. For every pair of UIOs in the bank, we compute the largest difference between their estimates and select the pair leading to the smallest difference. We prove that these observers provide exponential estimate of the system state. Once we have an estimate of the state, we provide tools for reconstructing attack signals using model matching techniques. Attacked actuators and sensors are isolated by simply checking the sparsity of the estimated attack signals. Finally, after obtaining state estimates and isolation has been performed, we provide a control scheme for stabilizing the closed-loop dynamics. In the case with sensor attacks only (no actuators attacks), we show that a separation principle between estimation and control holds and the system can be stabilized by closing the loop with the multi-observer estimator and a static output feedback controller. When both sensors and actuator are attacks, we propose an effective technique to stabilize the system by switching off the isolated actuators, and closing the loop with a multi-observer based output time-varying feedback controller. Because attack signals might be zero for some time instants, actuators isolated as attack-free might arbitrarily switch among all the supersets of the set of attack-free actuators. Therefore, we need a controller able to stabilize the closed-loop dynamics under the arbitrary switching induced by turning off the isolated actuators. To achieve this, we assume that a \textit{state} feedback controller that stabilizes the switching closed-loop system exists, and use this controller together with the multi-observer estimator to stabilize the system. We use Input-to-State Stability (ISS) \cite{Sontag2008} of the closed-loop system with respect to the exponentially stable estimation error to conclude on stability of the closed-loop dynamics. Compared to the adaptive controller proposed in \cite{Yadegar2017}, where a particular class of attacks is considered and ultimate boundedness of the closed-loop system is guaranteed only, our controller is able drive the system state asymptotically to the origin under arbitrary and potentially unbounded attack signals.\\[1mm]	
The paper is organized as follows. In Section 2, we present some preliminary results needed for the subsequent sections. In Section 3, we introduce the proposed UIO-based estimation schemes. In Section 4, a method for isolating actuator attacks is described. The proposed control scheme is given in Section \ref{control}. Finally, in Section 6, we give concluding remarks.

	\section{Preliminaries}
	\subsection{Notation}
	We denote the set of real numbers by $\mathbb{R}$, the set of natural numbers by $\mathbb{N}$ , the set of integers by $\mathbb{Z}$, and $\mathbb{R}^{n\times m}$ the set of $n\times m$ matrices for any $m,n \in \mathbb{N}$. For any vector $v\in\mathbb{R}^{n}$,  we denote {$v^{J}$} the stacking of all $v_{i}$, $i\in J$, $J\subset \left\lbrace 1,\hdots,n\right\rbrace$, $|v|=\sqrt{v^{\top} v}$, and $\supp(v)=\left\lbrace i\in\left\lbrace 1,\hdots,n\right\rbrace |v_{i}\neq0\right\rbrace $.  For matrices $C \in\mathbb{R}^{p \times n}$, $C^{\top} = (c_1^{\top},\ldots,c_p^{\top})$, we denote $C^J$ the stacking of all rows $c_{i} \in\mathbb{R}^{1 \times n}$, $i\in J$, $J\subset \left\lbrace 1,\hdots,n\right\rbrace$. Set $J$ is called a superset of set $S$ if $S\subseteq J$. We denote the cardinality of a set $S$ as $\card(S)$. The binomial coefficient is denoted as $\binom{a}{b}$, where $a,b$ are nonnegative integers. We denote a variable $m$ uniformly distributed in the interval $(z_{1},z_{2})$ as $m\sim\mathcal{U}(z_{1},z_{2})$ and normally distributed with mean $\mu$ and variance $\sigma^2$ as $m\sim \mathcal{N}(\mu,\sigma^2)$. The notation $\mathbf{0}_{n}$ and $I_{n}$ denote the zero matrix and the identity matrix of dimension $n \times n$, respectively. We simply write $\mathbf{0}$ and $I$ when their dimensions are evident. A continuous function $\alpha:[0,a)\to[0,\infty)$ is said to belong to class K, if it is strictly increasing and $\alpha(0)=0$, \cite{Khalil:1173048}. Similarity, a continuous function $\beta:[0,a)\times[0,\infty)\to[0,\infty)$ is said to belong to class KL if, for fixed $s$, the mapping $\beta(r,s)$ belongs to class K with respect to $r$ and, for fixed $r$, the mapping $\beta(r,s)$ is decreasing with respect to $s$ and $\beta(r,s)\to 0$ as $s\to\infty$, \cite{Khalil:1173048}.
	\section{Estimation}
	In \cite{Chong2015c}, the problem of state estimation for continuous-time LTI system under sensor attacks is solved using a bank of Luenberger observers. Inspired by these results, we use a bank of UIOs to estimate the state of the system when sensor and actuator attacks both occur.
	Consider a discrete-time linear system under sensor and actuator attacks:
	\begin{equation}\label{s1}
	\left\{ \begin{split}
	x^{+}=&Ax+B(u+a_{u})\\
	y=&Cx+a_{y}
	\end{split}\right.
	\end{equation}
	with state $x\in\mathbb{R}^{n}$, output $y\in\mathbb{R}^{n_{y}}$, known input $u\in\mathbb{R}^{n_{u}}$, vector of actuator attacks $a_{u}\in\mathbb{R}^{n_{u}}, a_{u}=(a_{u1},\ldots,a_{un_{u}})^{\top}$, i.e., $a_{ui}(k)=0$ for all $k\geq 0$ if the $i$-th actuator is attack-free; otherwise, $a_{ui}(k_{i})\neq 0$ for some $k_{i}\geq 0$ and can be arbitrarily large, and vector of sensor attacks $a_{y}\in\mathbb{R}^{n_{y}}, a_{y}=(a_{y1},\ldots,a_{yn_{y}})^{\top}$, i.e., $a_{yi}(k)=0$ for all $k\geq 0$ if the $i$-th sensor is attack-free; otherwise, $a_{yi}(k_{i})\neq 0$ for some $k_{i}\geq 0$ and can be arbitrarily large. Matrices $A,B,C$ are of appropriate dimensions, and we assume that $(A,B)$ is stabilizable, $(A,C)$ is detectable, and $B$ has full column rank. Let $W_{u}\subset\left\lbrace 1,\hdots,n_{u}\right\rbrace $ denotes the \emph{unknown} set of attacked actuators, and $W_{y}\subset\left\lbrace 1,\hdots,n_{y}\right\rbrace $ denotes the \emph{unknown} set of attacked sensors.
	\begin{assumption}\label{asump2}
		The sets of attacked actuators and sensors do not change over time, i.e., $W_{u}\subset\left\lbrace 1,\ldots,n_{u}\right\rbrace ,W_{y}\subset\left\lbrace 1,\ldots,n_{y}\right\rbrace $ are constant (time-invariant) and $\supp(a_{u}(k))\subseteq W_{u}$, $\supp(a_{y}(k))\subseteq W_{y}$, for all $k\geq 0$.
	\end{assumption}
	\subsection{Complete Unknown Input Observers}\label{complete}
We first treat $(u+a_{u})$ as an unknown input to system (\ref{s1}) and consider a UIO with the following structure:
	\begin{equation}\label{o}
	\left\{\begin{split}
	z_{J_s}^{+}=&N_{J_s}z_{J_s}+L_{J_s}y^{J_s},\\
	\hat{x}_{J_s}=&z_{J_s}+E_{J_{s}}y^{J_s},
	\end{split}\right.
	\end{equation}
	where $z_{J_{s}}\in\mathbb{R}^{n}$ is the state of the observer, $\hat{x}_{J_{s}}\in\mathbb{R}^{n}$ denotes the estimate of the system state, $(N_{J_s},L_{J_s},E_{J_s})$ are observer matrices of appropriate dimensions to be designed. It is easy to verify that if $(N_{J_s},L_{J_s},E_{J_s})$ satisfy the following equations:
	\begin{equation}\label{eq}
	\left\{\begin{split}
	N_{J_{s}}(I-E_{J_{s}}C^{J_{s}})+L_{J_{s}}C^{J_{s}}+(E_{J_{s}}C^{J_{s}}-I)A=&0,\\
	(E_{J_{s}}C^{J_{s}}-I)B=&0;
	\end{split}\right.
	\end{equation}
then, the estimation error $e_{J_s}=\hat{x}_{J_s}-x$ satisfies:
	\begin{equation}\label{error_Dyn1}
	e_{J_s}^{+}=N_{J_s}e_{J_s}.
	\end{equation}
If $N_{J_s}$ is Schur, system (\ref{o}) is called a UIO for (\ref{s1}). In \cite{Ding2013}, it is proved that such observer exists if and only if the following two conditions are satisfied:\vspace{1mm}

	\textbf{($\text{c}_\text{1}$)} $\rank(C^{J_s}B)=\rank(B)=n_{u}$.\vspace{1mm}
	
	\textbf{($\text{c}_\text{2}$)} The pair $(C^{J_s},A-E_{J_s}C^{J_s}A)$ is detectable.\vspace{1mm}
	
\begin{figure}[t]\centering
		\includegraphics[width=0.45\textwidth]{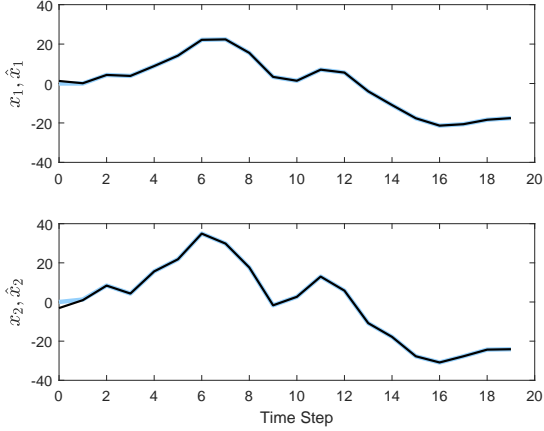}
		\caption{Estimated states $\hat{x}$ converges to the true states $x$ when $a_{u},a_{y3}\sim\mathcal{U}(-10,10)$. Legend: $\hat{x}$ (blue), true states (black)}
		\label{fig:2e}
		\centering
	\end{figure}

Let $q$ be the largest integer such that for all $J_{s}\subset\left\lbrace 1,\ldots,n_{y}\right\rbrace $ with $\card(J_{s})\geq n_{y}-2q>0$, conditions $(c_{1})$ and $(c_{2})$ are satisfied;
	then, observer (\ref{o}) can be constructed for any $C^{J_{s}}$ with $\card(J_{s})\geq n_{y}-2q$ by solving (\ref{eq}) for a Schur matrix $N_{J_s}$. Hence, for such an observer, if $a_{y}^{J_{s}}(k)=0$ for all $k\geq 0$, there exist $c_{J_s}>0$, $\lambda_{J_s}\in(0,1)$ satisfying:
	\begin{equation}
	|e_{J_s}(k)|\leq c_{J_s}\lambda_{J_s}^{k}|e_{J_s}(0)|,
	\end{equation}
	for all $k\geq 0$ \cite{Ding2013}, where $e_{J_{s}}=\hat{x}_{J_{s}}-x$.
	\begin{assumption}\label{a1}
		There are at most $q$ sensors attacked by an adversary, i.e.,
		\begin{equation}
		\card(W_{y})\leq q<\frac{n_{y}}{2},
		\end{equation}
		where $q$ is the largest positive integer satisfying conditions $(c_{1})$ and $(c_{2})$.
	\end{assumption}
	\begin{lemma}\label{l4}
		Under Assumption \ref{a1}, among each set of $n_{y}-q$ sensors, at least $n_{y}-2q>0$ of them are attack-free.
	\end{lemma}
	\textbf{\emph{Proof:}}
	Lemma \ref{l4} follows trivially from Assumption \ref{a1}. \hfill$\blacksquare$\\[1mm]
	Let Assumption \ref{a1} be satisfied. Inspired by the ideas in \cite{Chong2015}, we use a UIO for each subset $J_{s}\subset\left\lbrace 1,\ldots,n_{y}\right\rbrace $ of sensors with $\card(J_{s})=n_{y}-q$ and for each subset $S_{s}\subset\left\lbrace 1,\ldots,n_{y}\right\rbrace $ of sensors with $\card(S_{s})=n_{y}-2q$. Under Assumption \ref{a1}, there exists at least one set $\bar{J}_{s}\subset\left\lbrace 1,\ldots,n_{y}\right\rbrace $ with $\card(\bar{J}_{s})=n_{y}-q$ such that $a_{y}^{J_{s}}(k)=0$ for all $k\geq 0$. Then, the estimate given by the UIO for $\bar{J}_{s}$ is a correct estimate, and the estimate given by the UIO for any $S_{s}\subset\bar{J}_{s}$ with $\card(S_{s})=n_{y}-2q$ is consistent with that given by $\bar{J}_{s}$. This motivates the following estimation strategy. \\[1mm]
For each set $J_{s}$ with $\card(J_{s})=n_{y}-q$, we define $\pi_{J_{s}}(k)$ as the largest deviation between $\hat{x}_{J_{s}}$ and $\hat{x}_{S_{s}}$ that is given by any $S_{s}\subset J_{s}$ with $\card(S_{s})=n_{y}-2q$, i.e.,
	\begin{equation}\label{es1}
	\pi_{J_{s}}(k):=\max_{S_{s}\subset J_{s}:\card(S_{s})=n_{y}-2q}|\hat{x}_{J_{s}}(k)-\hat{x}_{S_{s}}(k)|,
	\end{equation}
for all $k\geq 0$, and the sequence $\sigma_{s}(k)$ as
	\begin{equation}\label{es2}
	\sigma_{s}(k):=\argmin_{J_{s}\subset\left\lbrace 1,\ldots,n_{y}\right\rbrace :\card(J_{s})=n_{y}-q}\pi_{J_{s}}(k).
	\end{equation}
	Then, as proved below, the estimate indexed by $\sigma_{s}(k)$:
	\begin{equation}\label{es3}
	\hat{x}(k):=\hat{x}_{\sigma_{s}(k)}(k),
	\end{equation}
is an exponential attack-free estimate of the system
state. For simplicity and without generality, for all $J_{s}$ and $S_{s}$, $z_{J_{s}}(0)$ and $z_{S_{s}}(0)$ are chosen such that $\hat{x}_{J_{s}}(0)=\hat{x}_{S_{s}}(0)=\hat{x}(0)$. The following result summarizes the ideas presented above.
	\begin{theorem}\label{th1}
		Consider system \eqref{s1}, observer \eqref{o}, and the complete multi-observer estimator \eqref{es1}-\eqref{es3}. Define the estimation error $e(k):=\hat{x}_{\sigma_{s}(k)}(k)-x(k)$, and let conditions $(c_{1})$-$(c_{2})$ and Assumptions \ref{asump2}-\ref{a1} be satisfied; then, there exist constants $\bar{c}>0$, $\bar{\lambda}\in(0,1)$ satisfying:
		\begin{equation}
		|e(k)|\leq\bar{c}\bar{\lambda}^{k}|e(0)|\label{sa},
		\end{equation}
		for all $e(0)\in\mathbb{R}^{n}$, $k\geq 0$.
	\end{theorem}
	\emph{\textbf{Proof:}}
	Under Assumption \ref{a1}, there exists at least one set $\bar{J}_{s}$ with $\card(\bar{J}_{s})=n_{y}-q$ such that $a_{y}^{\bar{J}_{s}}(k)=0$ for all $k\geq 0$. Then, there exist $c_{\bar{J}_{s}}>0$ and $\lambda_{\bar{J}_{s}}\in(0,1)$ such that
	\begin{equation}\label{6}
	|e_{\bar{J}_{s}}(k)|\leq c_{\bar{J}_{s}}\lambda_{\bar{J}_{s}}^{k}|e(0)|,
	\end{equation}
	for all $e(0)\in\mathbb{R}^{n}$ and $k\geq0$. Moreover, for any set $S_{s}\subset\bar{J}_{s}$ with $\card(S_{s})=n_{y}-2q$, we have $a_{y}^{S_{s}}(k)=0$ $\forall k\geq0$; hence, there exist $c_{S_{s}}>0$ and $\lambda_{S_{s}}\in(0,1)$ such that
	\begin{equation}\label{7}
	|e_{S_{s}}(k)|\leq c_{S_{s}}\lambda_{S_{s}}^{k}|e(0)|,
	\end{equation}
	for all $e(0)\in\mathbb{R}^{n}$ and $k\geq 0$.  Consider $\pi_{J_{s}}$ in (\ref{es1}). Combining the above inequalities, we have
	\begin{equation}\label{sv}
	\begin{split}
	\pi_{\bar{J}_{s}}(k)=&\underset{S_{s}\subset\bar{J}_{s}}{\max}|\hat{x}_{\bar{J}_{s}}(k)-\hat{x}_{S_{s}}(k)|\\
	=&\underset{S_{s}\subset\bar{J}_{s}}{\max}|\hat{x}_{\bar{J}_{s}}(k)-x(k)+x(k)-\hat{x}_{S_{s}}(k)|\\
	\leq&  |e_{\bar{J}_{s}}(k)|+\underset{S_{s}\subset\bar{J}_{s}}{\max}|e_{S_{s}}(k)|\\
	\leq& 2c'_{\bar{J}_{s}}\lambda_{\bar{J}_{s}}^{'k}|e(0)|,
	\end{split}
	\end{equation}
	for all $e(0)\in\mathbb{R}^{n}$ and $k\geq 0$, where \[c'_{\bar{J}_{s}}:=\underset{S_{s}\subset\bar{J}_{s}}{\max}\left\lbrace c_{\bar{J}_{s}}, c_{S_{s}}\right\rbrace,\] \[\lambda'_{\bar{J}_{s}}:=\underset{S_{s}\subset\bar{J}_{s}}{\max}\left\lbrace \lambda_{\bar{J}_{s}}, \lambda_{S_{s}}\right\rbrace.\]
	Note that $S_{s}\subset \bar{J}_{s}$, $\card(S_{s})=n_{y}-2q_{2}$. Then, from (\ref{es2}), we have $\pi_{\sigma_{s}(k)}(k)\leq\pi_{\bar{J}_{s}}(k)$. From Lemma \ref{l4}, we know that there exist at least one set $\bar{S}_{s}\subset\sigma_{s}(k)$ with $\card(\bar{S}_{s})=n_{y}-2q$, such that $a_{y}^{\bar{S}_{s}}(k)=0$ for all $k\geq 0$, and there exist $c_{\bar{S}_{s}}>0$ and $\lambda_{\bar{S}_{s}}\in(0,1)$ such that
	\begin{equation}\label{ss}
	|e_{\bar{S}_{s}}(k)|\leq c_{\bar{S}_{s}}\lambda_{\bar{S}_{s}}^{k}|e(0)|,
	\end{equation}
	for all $e(0)\in\mathbb{R}^{n}$ and $k\geq0$. From (\ref{es1}), we have \[
	\begin{split}
	\pi_{\sigma_{s}(k)}(k)=&\underset{S_{s}\subset\sigma_{s}(k)}{\max}|\hat{x}_{\sigma_{s}(k)}(k)-\hat{x}_{S_{s}}(k)|\\\geq&|\hat{x}_{\sigma_{s}(k)}(k)-\hat{x}_{\bar{S}_{s}}(k)|.
	\end{split}
	\] Using this lower bound on $\pi_{\sigma_{s}(k)}(k)$ and the triangle inequality we have that
	\begin{equation}
	\begin{split}
	|e_{\sigma_{s}(k)}(k)|=&|\hat{x}_{\sigma_{s}(k)}(k)-x(k)|\\
	=&|\hat{x}_{\sigma_{s}(k)}(k)-\hat{x}_{\bar{S}_{s}}(k)+\hat{x}_{\bar{S}_{s}}(k)-x(k)|\\
	\leq&|\hat{x}_{\sigma_{s}(k)}(k)-\hat{x}_{\bar{S}_{s}}(k)|+|e_{\bar{S}_{s}}(k)|\\
	\leq&\pi_{\sigma_{s}(k)}(k)+|e_{\bar{S}_{s}}(k)|\\
	\leq&\pi_{\bar{J}_{s}}(k)+|e_{\bar{S}_{s}}(k)|,
	\end{split}
	\end{equation}
	for all $k\geq 0$. Hence, from (\ref{sv}) and (\ref{ss}), we have
	\begin{equation}\label{sb}
	|e_{\sigma_{s}(k)}(k)|\leq \bar{c}\bar{\lambda}^{k}|e(0)|,
	\end{equation}
	for all $e(0)\in\mathbb{R}^{n}$ and $k\geq 0$, where $\bar{c}=3\max\{ c_{\bar{S}_{s}},c'_{\bar{J}_{s}}\} $ and $\bar{\lambda}=\max\{ \lambda_{\bar{S}_{s}},\lambda'_{\bar{J}_{s}}\} $. Inequality (\ref{sb}) is of the form (\ref{sa}) and the result follows.\hfill$\blacksquare$\\[1mm]
	\textbf{Example 1:}
	Consider the following system subject to actuator and sensor attacks:
	\begin{equation}
	\left\{\begin{split}\label{e1}
	x^{+}=&\left[ \begin{matrix}
	0.2&0.5\\
	0.2&0.7
	\end{matrix}\right] x+\left[ \begin{matrix}
	1\\
	2\\
	\end{matrix}\right] (u+a_{u}),\\
	y=&\left[ \begin{matrix}
	1&3\\
	1&1\\
	3&2\\
	2&1
	\end{matrix}\right] x+a_{y}.
	\end{split}\right.
	\end{equation}
It can be verified that a UIO of the form (\ref{o}) exists for each $C^{J_{s}}$ with  $J_{s}\subset\left\lbrace 1,2,3,4\right\rbrace $ and $\card(J_{s})\geq 2$; then, $4-2q=2$, i.e., $q=1$ and at most one sensor is attacked. We attack the actuator and let $W_{y}=\left\lbrace 3\right\rbrace $, i.e., the third sensor is attacked. We let $u\sim\mathcal{U}(-1,1)$, $a_{u},a_{y3}\sim\mathcal{U}(-10,10)$. We design a UIO for each $J_{s}$ with $\card(J_{s})=3$, and for each $S_{s}$ with $\card(S_{s})=2$. Therefore, totally $\binom{4}{3}+\binom{4}{2}=10$ UIOs are designed and they are all initialized at $\hat{x}(0)=\left[ 0,0\right] ^{\top}$. For $k\in[0,19]$, the estimator \eqref{o}, (\ref{54})-(\ref{56}) is used to construct $\hat{x}(k)$. The performance of the estimator is shown in Figure \ref{fig:2e}.
\subsection{Partial Unknown Input Observers}\label{partial}
	Here, we are implicitly assuming that either condition $(c_{1})$ or $(c_{2})$ (or both) cannot be satisfied for any $C^{J_{s}}$ with $\card(J_{s})=n_{y}-2q$ with $q\geq 1$. Let $B$ be partitioned as $B=\left[ b_{1},\hdots,b_{i},\hdots,b_{n_{u}}\right] $ where $b_{i}\in\mathbb{R}^{n\times 1}$ is the $i$-th column of $B$. Then, the attacked system (\ref{s1}) can be written as
	\begin{equation}\label{s2}
	\left\{\begin{split}
	x^{+}=&Ax+Bu+b_{W_{u}}a^{W_{u}},  \\
	y=&Cx+a_{y},
	\end{split}\right.
	\end{equation}
	where the attack input $a^{W_{u}}$ can be regarded as an unknown input and the columns of $b_{W_{u}}$ are $b_{i}$, $i\in W_{u}$. Denote by $b_{J_{u}}$ the matrix whose columns are $b_{i}$ for $i\in J_{u}$. Let $q_{1}$ and $q_{2}$ be the largest integers such that for all $J_{u}\subset\left\lbrace 1,\ldots,n_{u}\right\rbrace $ with $\card(J_{u})\leq 2q_{1}<n_{u}$ and $J_{s}\subset\left\lbrace 1,\ldots,n_{y}\right\rbrace $ with $\card(J_{s})\geq n_{y}-2q_{2}>0$, the following is satisfied: \vspace{1mm}
	
	\textbf{($\text{c}_\text{3}$)} $\rank(C^{J_{s}}b_{J_{u}})=\rank(b_{J_{u}})=\card(J_{u}).$\vspace{1mm}
	
	\textbf{($\text{c}_\text{4}$)} There exists $(N_{J_{us}},L_{J_{us}},E_{J_{us}},T_{J_{us}})$ satisfying:
	\begin{equation}\label{eq2}
	\left\{\begin{split}
	N_{J_{us}}(I-E_{J_{us}}C^{J_{s}})+L_{J_{us}}C^{J_{s}}+(E_{J_{us}}C^{J_{s}}-I)A=&0,\\
	(T_{J_{us}}+E_{J_{us}}C^{J_{s}}-I)B=&0,\\
	(E_{J_{us}}C^{J_{s}}-I)b_{J_{u}}=&0,
	\end{split}\right.
	\end{equation}
with detectable pair $(C^{J_{us}},A-E_{J_{us}}C^{J_{us}}A)$ and Schur $N_{J_{us}}$. If conditions $(c_{3})$ and $(c_{4})$ are satisfied, a UIO with the following structure exists for each $b_{J_{u}}$ with $J_{u}\subset \left\lbrace 1,\hdots,n_{u}\right\rbrace$, $\card(J_{u})\leq 2q_{1}<n_{u}$ and each $C^{J_{s}}$ with $J_{s}\subset\left\lbrace 1,\ldots,n_{y}\right\rbrace$, $\card(J_{s})\geq n_{y}-2q_{2}>0$:
	\begin{equation}
	\left\{\begin{split}\label{o1}
	z_{J_{us}}^{+}=&N_{J_{us}}z_{J_{us}}+T_{J_{us}}Bu+L_{J_{us}}y^{J_{s}},\\
	\hat{x}_{J_{us}}=&z_{J_{us}}+E_{J_{us}}y^{J_{s}},
	\end{split}\right.
	\end{equation}
	where $z_{J_{us}}\in\mathbb{R}^{n}$ is the observer state, $\hat{x}_{J_{us}}$ denotes the state estimate, and $(N_{J_{us}},L_{J_{us}},T_{J_{us}},E_{J_{us}})$ are the observer matrices satisfying (\ref{eq2}), see \cite{Ding2013} for further details. That is, system (\ref{o1}) is a UIO for the system:
	\begin{equation}\label{s3}
	\left\{\begin{split}
	x^{+}=&Ax+Bu+b_{J_{u}}a_{u}^{J_{u}},  \\
	y^{J_{s}}=&C^{J_{s}}x+a_{y}^{J_{s}},
	\end{split}\right.
	\end{equation}
	with unknown input $b_{J_{u}}a^{J_{u}}$ and known input $Bu$. It follows that the estimation error $e_{J_{us}}=\hat{x}_{J_{us}}-x$ satisfies:
	\begin{equation}
	e_{J_{us}}^{+}=N_{J_{us}}e_{J_{us}},
	\end{equation}
for some Schur matrix $N_{J_{us}}$. We refer to UIOs of the form (\ref{s3}) as \emph{partial} UIOs for the pair $(J_{u},J_{s})$.
	\begin{assumption}\label{asump1}
		There are at most $q_{1}$ actuators and at most $q_{2}$ sensors attacked by an adversary, i.e.,
		\begin{equation}
		\card(W_{u})\leq q_{1}<\frac{n_{u}}{2}
		\end{equation}
		\begin{equation}
		\card(W_{y})\leq q_{2}<\frac{n_{y}}{2},
		\end{equation}
		where $q_{1}$ and $q_{2}$ are the largest positive integers satisfying $(c_{3})$ and $(c_{4})$.
	\end{assumption}
	\begin{remark}
	Note that if conditions $(c_{3})$ and $(c_{4})$ are satisfied for $b_{J_{u}}$ with $\card(J_{u})=2q_{1}=n_{u}$, then conditions $(c_{1})$ and $(c_{2})$ are satisfied, and \eqref{o1} is a complete UIO for \eqref{s1} for $T_{J_{us}}=\mathbf{0}$. Since we are considering partial UIOs, we assume $2q_{1}<n_{u}$ to exclude this case.
	\end{remark}
\begin{lemma}\label{lm2}
Under Assumption \ref{asump1}, for each set of $q_{1}$ actuators, among all its supersets with $2q_{1}$ actuators, at least one set is a superset of $W_{u}$.
\end{lemma}
	\begin{lemma}\label{lm4}
		Under Assumption \ref{asump1}, among each set of $n_{y}-q_{2}$ sensors, at least $n_{y}-2q_{2}>0$ sensors are attack-free.
	\end{lemma}
	\emph{\textbf{Proof:}}
Lemmas \ref{lm2} and \ref{lm4} follow trivially from Assumption \ref{asump1}. \hfill$\blacksquare$
	
\begin{figure}[t]\centering
		\includegraphics[width=0.45\textwidth]{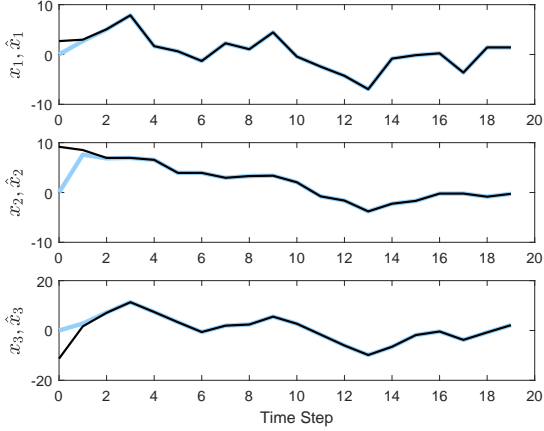}
		\caption{Estimated states $\hat{x}$ converges to the true states $x$ when $a_{u3},a_{y2}\sim\mathcal{U}(-10,10)$. Legend: $\hat{x}$ (blue), true states (black)}
		\label{fig:2}
		\centering
	\end{figure}

Note that the existence of a UIO for each pair $(J_{u},J_{s})$ with $\card(J_{u})\leq 2q_{1}$ and $\card(J_{s})\geq n_{y}-2q_{2}$ means that if $W_{u}\subseteq J_{u}$ and $a_{y}^{J_{s}}(k)=0$ for all $k\geq 0$, the estimation error $e_{J_{us}}=\hat{x}_{J_{us}}-x$ satisfies
	\begin{equation}
	|e_{J_{us}}|\leq c_{J_{us}}\lambda_{J_{us}}^{k}|e_{J_{us}}(0)|,
	\end{equation}
for some $c_{J_{us}}>0$ and $\lambda_{J_{us}}\in(0,1)$, all $e_{J_{us}}(0)\in\mathbb{R}^{n}$, and $k\geq 0$. Let Assumption \ref{asump1} be satisfied. We use a UIO for each pair $(J_{u},J_{s})$ with $\card(J_{u})=q_{1}$ and $\card(J_{s})=n_{y}-q_{2}$. Then, we use a UIO for each pair $(S_{u},S_{s})$ with $S_{u}\subset\left\lbrace 1,\ldots,n_{u}\right\rbrace$, $\card(S_{u})=2q_{1}$ and $S_{s}\subset\left\lbrace 1,\ldots,n_{y}\right\rbrace$, $\card(S_{s})=n_{y}-2q_{2}$. Under Assumption \ref{asump1} , there exists at least one set $\bar{J}_{u}$ with $\card(\bar{J}_{u})=q_{1}$ such that $W_{u}\subseteq\bar{J}_{u}$ and at least one set $\bar{J}_{s}$ with $\card(\bar{J}_{s})=n_{y}-q_{2}$ such that $a_{y}^{\bar{J}_{s}}(k)=0$ for all $k\geq 0$. Then, the estimate given by the UIO for $(\bar{J}_{u},\bar{J}_{s})$ is a correct estimate, and the estimates given by the UIOs for any $(S_{u},S_{s})$ (denoted as $\hat{x}_{S_{us}}$), where $S_{u}\supset\bar{J}_{u}$, $\card(S_{u})=2q_{1}$ and $S_{s}\subset \bar{J}_{s}$, $\card(J_{s})=n_{y}-2q_{2}$, are consistent with $\hat{x}_{J_{us}}$. This motivates the following estimation strategy.\\[1mm]
For each pair $(J_{u},J_{s})$ with $\card(J_{u})=q_{1}$ and $\card(J_{s})=n_{y}-q_{2}$, define $\pi_{J_{us}}(k)$ as the largest deviation between $\hat{x}_{J_{us}}(k)$ and $\hat{x}_{S_{us}}(k)$ that is given by any pair $(S_{u},S_{s})$, where $S_{u}\supset J_{u}$ with $\card(S_{u})=2q_{1}$ and  $S_{s}\subset J_{s}$ with $\card(S_{s})=n_{y}-2q_{2}$. That is,
	\begin{equation} \label{54}
	\pi_{J_{us}}(k):=\max_{S_{u}\supset J_{u},S_{s}\subset J_{s}}|\hat{x}_{J_{us}}(k)-\hat{x}_{S_{us}}(k)|,
	\end{equation}
	for all $k\geq 0$. Define the sequences $\sigma_{u}(k)$ and $\sigma_{s}(k)$ as
	\begin{equation}\label{55}
	(\sigma_{u}(k),\sigma_{s}(k)):=\underset{J_{u},J_{s}}{\argmin}\hspace{2mm}\pi_{J_{us}}(k).
	\end{equation}
Then, as proven below, the estimate indexed by ($\sigma_{u}(k),\sigma_{s}(k)$):
	\begin{equation}\label{56}
	\hat{x}(k)=\hat{x}_{\sigma_{us}(k)}(k),
	\end{equation}
is an exponential attack-free estimate of the system state. For simplicity and without generality, for all $J$ and $S$, $z_{J_{us}}(0)$ and $z_{S_{us}}(0)$ are chosen such that $\hat{x}_{J_{us}}(0)=\hat{x}_{S_{us}}(0)=\hat{x}(0)$. The following result summarizes the ideas presented above.
	\begin{theorem}\label{t1}
		Consider system \eqref{s1}, observer \eqref{o1}, and the partial multi-observer estimator \eqref{54}-\eqref{56}. Define the estimation error $e(k):=\hat{x}_{\sigma_{us}(k)}(k)-x(k)$ and let $(c_{3})$-$(c_{4})$ and Assumptions \ref{asump2},\ref{asump1} be satisfied; then, there exist positive constants $\bar{c}>0$ and $\bar{\lambda}\in(0,1)$ satisfying:
		\begin{equation}
		|e(k)|\leq\bar{c}\bar{\lambda}^{k}|e(0)|\label{60},
		\end{equation}
		for all $e(0)\in\mathbb{R}^{n}$, $k\geq 0$.
	\end{theorem}
	\emph{\textbf{Proof:}}
	Under Assumption \ref{asump1}, there exists at least one set $\bar{J}_{u}$ with $\card(\bar{J})=q_{1}$ such that $\bar{J}_{u}\supset W_{u}$, and at least one set $\bar{J}_{s}$ with $\card(\bar{J}_{s})=n_{y}-q_{2}$ such that $a_{y}^{\bar{J}_{s}}(k)=0$ for all $k\geq 0$; then, there exist $c_{\bar{J}_{us}}>0$ and $\lambda_{\bar{J}_{us}}\in(0,1)$ satisfying
	\begin{equation}\label{63}
	|e_{\bar{J}_{us}}(k)|\leq c_{\bar{J}_{us}}\lambda_{\bar{J}_{us}}^{k}|e(0)|,
	\end{equation}
	for all $e(0)\in\mathbb{R}^{n}$ and $k\geq0$. Moreover, for any set $S_{u}\supset\bar{J}_{u}$ with $\card(S_{u})=2q_{1}$ and $S_{s}\subset\bar{J}_{s}$ with $\card(S_{s})=n_{y}-2q_{2}$, we have $S_{u}\supset W_{u}$ and $a_{y}^{S_{s}}(k)=0$ for all $k\geq0$; hence, there exist $c_{S_{us}}>0$ and $\lambda_{S_{us}}\in(0,1)$ such that
	\begin{equation}\label{61}
	|e_{S_{us}}(k)|\leq c_{S_{us}}\lambda_{S_{us}}^{k}|e(0)|,
	\end{equation}
	for all $e(0)\in\mathbb{R}^{n}$ and $k\geq 0$. Consider $\pi_{\bar{J}_{us}}$ in (\ref{54}). Combining the above results, we have that
	\begin{eqnarray*}
		\begin{split}
			\pi_{\bar{J}_{us}}(k)&=\underset{S_{u}\supset\bar{J}_{u},S_{s}\subset\bar{J}_{s}}{\max}|\hat{x}_{\bar{J}_{us}}(k)-\hat{x}_{S_{us}}(k)|\\
			&=\underset{S_{u}\supset\bar{J}_{u},S_{s}\subset\bar{J}_{s}}{\max}|\hat{x}_{\bar{J}_{us}}(k)-x(k)+x(k)-\hat{x}_{S_{us}}(k)|\\
			&\leq  |e_{\bar{J}_{us}}(k)|+\underset{S_{u}\supset\bar{J}_{u},S_{s}\subset\bar{J}_{s}}{\max}|e_{S_{us}}(k)|,
		\end{split}
	\end{eqnarray*}
	for all $k\geq 0$. From (\ref{63}) and (\ref{61}), we obtain
	\begin{equation}\label{66}
	\pi_{\bar{J}_{us}}(k)\leq 2c'_{\bar{J}_{us}}\lambda_{\bar{J}_{us}}^{'k}|e(0)|,
	\end{equation}
	for all $e(0)\in\mathbb{R}^{n}$ and $k\geq 0$, where \[c'_{\bar{J}_{us}}:=\underset{S_{u}\supset\bar{J}_{u},S_{s}\subset\bar{J}_{s}}{\max}\left\lbrace c_{\bar{J}_{us}}, c_{S_{us}}\right\rbrace,\] \[\lambda'_{\bar{J}_{us}}:=\underset{S_{u}\supset\bar{J}_{u},S_{s}\subset\bar{J}_{s}}{\max}\left\lbrace \lambda_{\bar{J}_{us}}, \lambda_{S_{us}}\right\rbrace.\]
	Note that $S_{u}\supset \bar{J}_{u}$, $\card(J_{u})=2q_{1}$, and $S_{s}\subset \bar{J}_{s}$, $\card(S_{s})=n_{y}-2q_{2}$. Then, from (\ref{55}), we have $\pi_{\sigma_{us}(k)}(k)\leq\pi_{\bar{J}_{us}}(k)$. By Lemmas \ref{lm2} and \ref{lm4}, we know that there exists at least one set $\bar{S}_{u}\supset\sigma_{u}(k)$ with $\card(\bar{S}_{u})=2q_{1}$ and at least one set $\bar{S}_{s}\subset\sigma_{s}(k)$ with $\card(\bar{S}_{s})=n_{y}-2q_{1}$ such that $\bar{S}_{u}\supset W_{u}$ and $a_{y}^{\bar{S}_{s}}(k)=0$ for all $k\geq 0$. Hence, there exist $c_{\bar{S}_{us}}>0$ and $\lambda_{\bar{S}_{us}}\in(0,1)$ satisfying
	\begin{equation}\label{67}
	|e_{\bar{S}_{us}}(k)|\leq c_{\bar{S}_{us}}\lambda_{\bar{S}_{us}}^{k}|e(0)|,
	\end{equation}
	for all $e(0)\in\mathbb{R}^{n}$ and $k\geq0$. From (\ref{54}), by construction \[
	\begin{split}
	\pi_{\sigma_{us}(k)}(k)=&\underset{S_{u}\supset \sigma_{u}(k),S_{s}\subset\sigma_{s}(k)}{\max}|\hat{x}_{\sigma_{us}(k)}(k)-\hat{x}_{S_{us}}(k)|\\\geq&|\hat{x}_{\sigma_{us}(k)}(k)-\hat{x}_{\bar{S}_{us}}(k)|.
	\end{split}
	\]
Using the above lower bound on $\pi_{\sigma_{us}(k)}(k)$ and the triangle inequality, we have that
	\begin{equation}
	\begin{split}
	|e_{\sigma_{us}(k)}(k)|=&|\hat{x}_{\sigma_{us}(k)}(k)-x(k)|\\
	=&|\hat{x}_{\sigma_{us}(k)}(k)-\hat{x}_{\bar{S}_{us}}(k)+\hat{x}_{\bar{S}_{us}}(k)-x(k)|\\
	\leq&|\hat{x}_{\sigma_{us}(k)}(k)-\hat{x}_{\bar{S}_{us}}(k)|+|e_{\bar{S}_{us}}(k)|\\
	\leq&\pi_{\sigma_{us}(k)}(k)+|e_{\bar{S}_{us}}(k)|\\
	\leq&\pi_{\bar{J}_{us}}(k)+|e_{\bar{S}_{us}}(k)|,
	\end{split}
	\end{equation}
	for all $k\geq 0$. Hence, from (\ref{66}) and (\ref{67}), we have
	\begin{equation}\label{70}
	|e_{\sigma_{us}(k)}(k)|\leq \bar{c}\bar{\lambda}^{k}|e(0)|,
	\end{equation}
	for all $e(0)\in\mathbb{R}^{n}$ and $k\geq 0$, where $\bar{c}=3\max\{ c_{\bar{S}_{us}},c'_{\bar{J}_{us}}\}$, $\bar{\lambda}=\max\{ \lambda_{\bar{S}_{us}},\lambda'_{\bar{J}_{us}}\}$. Inequality (\ref{70}) is of the form (\ref{60}), and the result follows.\hfill$\blacksquare$\\[1mm]	
	\textbf{Example 2:}
	Consider a linear system subject to actuator and sensor attacks:
	\begin{equation}
	\left\{\begin{split}
	x^{+}=&\left[ \begin{matrix}\label{e2}
	0.5&0&0.1\\
	0.2&0.7&0\\
	1&0&0.3
	\end{matrix}\right] x+\left[ \begin{matrix}
	0.5&0&0.5\\
	1&1&0.1\\
	0&0&0.5
	\end{matrix}\right] (u+a_{u}),\\
	y=&\left[ \begin{matrix}
	1&2&0\\0&1&1\\0&1&2\\1&1&1
	\end{matrix}\right] x+a_{y}.
	\end{split}\right.
	\end{equation}
It can be verified that complete UIOs do not exist for any $C^{J_{s}}$ with $\card(J_{s})\leq 2$. However, a partial UIO exists for each pair $(J_{u}, J_{s})$ with $\card(J_{u})\leq 2$ and $\card(J_{s})\geq 2$; then, $2q_{1}=2$ and $4-2q_{2}=2$, i.e., $q_{1}=q_{2}=1$. We let $W_{u}=\left\lbrace 3\right\rbrace $, $W_{y}=\left\lbrace 2\right\rbrace $, i.e., the third actuator and the second sensor are attacked, $u\sim\mathcal{U}(-1,1)$, and $a_{u3},a_{y2}\sim\mathcal{U}(-10,10)$. We construct a partial UIO for each pair $(J_{u},J_{s})$ with $\card(J_{u})=1,\card(J_{s})=3$ and each set $(S_{u},S_{s})$ with $\card(S_{u})=2,\card(S_{s})=2$. Therefore, totally $\binom{3}{1}\times\binom{4}{3}+\binom{3}{2}\times\binom{4}{2}=30$ partial UIOs are designed. We initiate the observers at $\hat{x}(0)=[0,0,0]^{\top}$. Estimator \eqref{o1}, (\ref{54})-(\ref{56}) is used to construct $\hat{x}(k)$. The performance of the estimator is shown in Figure \ref{fig:2}.
	\section{Attack Isolation and Reconstruction}
	Once we have an estimate $\hat{x}(k)$ of $x(k)$, either using the complete multi-observer estimator in Section \ref{complete} or the partial multi-observer estimator in Section \ref{partial}, we can use these estimates, the system model \eqref{s1}, and the known inputs to exponentially reconstruct the attack signals. Note that $e = \hat{x} - x \Rightarrow x = \hat{x} - e \Rightarrow x^{+} = \hat{x}^{+} - e^{+}$. Then, the system dynamics \eqref{s1} can be written in terms of $e$ and $\hat{x}$ as follows:
	\begin{equation}\label{attack_estimation_1}
	\left\{
	\begin{split}
	\hat{x}^{+} &= e^{+} + A(\hat{x} - e) + B(u+a_{u}),\\
	&\hspace{25mm}\Downarrow\\
	a_{u} &= B_{left}^{-1}(\hat{x}^{+}  -A\hat{x})-u - B_{left}^{-1}(e^+ - Ae), \\
	\end{split}
	\right.
	\end{equation}
	because $B$ has full column rank (as introduced in the system description), where $B_{left}^{-1}$ denotes the Moore-Penrose pseudoinverse of $B$. Similarly, we have
	\begin{equation}\label{attack_estimation_2}
	\left\{
	\begin{split}
	y&= Cx+a_{y}=C\hat{x}-Ce+a_{y},\\
	&\hspace{25mm}\Downarrow\\
	a_{y} &= y-C\hat{x}+Ce. \\
	\end{split}
	\right.
	\end{equation}
	First, consider the complete multi-observer in Section \ref{complete}. Let the estimation error dynamics characterized by \eqref{es1}-\eqref{es3} be given by
	\begin{equation}\label{he}
	e^{+}=f_{1}(e,x,a_{y},a_{u}),
	\end{equation}
	where $f_{1}:\mathbb{R}^n \times \mathbb{R}^n \times \mathbb{R}^{n_{y}}\times\mathbb{R}^{n_{u}} \rightarrow \mathbb{R}^n$ denotes some nonlinear function. That is, the estimation error is given by some nonlinear function of the state and the attack signals. However, in Theorem \ref{th1}, we have proved that $e$ converges to the origin exponentially. Hence, the terms depending on $e$ and $e^+$ in the expression for $a_{u}$ and $a_{y}$ in \eqref{attack_estimation_1} and \eqref{attack_estimation_2} vanishes exponentially and therefore, the following attack estimate:
	\begin{equation}\label{ru}
	\hat{a}_{u}(k) = B_{left}^{-1}(\hat{x}(k) - A\hat{x}(k-1))-u(k-1),
	\end{equation}
	and
	\begin{equation}\label{ry}
	\hat{a}_{y}(k)=y(k)-C\hat{x}(k),
	\end{equation}
	exponentially reconstruct the attack signals $a_{u}(k-1)$ and $a_{y}(k)$, i.e.,
	\begin{equation}
	\begin{split}
	\lim_{k\to\infty}(\hat{a}_{u}(k)-a_{u}(k-1))=0,
	\end{split}
	\end{equation}
	and
	\begin{equation}
	\lim_{k\to\infty}(\hat{a}_{y}(k)-a_{y}(k))=0.
	\end{equation}
	Then, for sufficiently large $k$, the sparsity pattern of $\hat{a}_{u}(k)$ and $\hat{a}_{y}(k)$ can be used to isolate attacks, i.e.,
	\begin{equation}\label{wu}
	\hat{W}_{u}(k)=\supp(\hat{a}_{u}(k)),
	\end{equation}
	and
	\begin{equation}\label{wy}
	\hat{W}_{y}(k)=\supp(\hat{a}_{y}(k)),
	\end{equation}
	where $\hat{W}_{u}(k)$ denotes the set of isolated attacked actuators, and $\hat{W}_{y}(k)$ denotes the set of isolated attacked sensors. Note that we can only estimate $a_{u}$ from $\hat{x}^{+}$ and $e^{+}$, which implies that we always have, at least, one-step delay for actuator attacks isolation.
	
\begin{figure}[t]\centering
		\includegraphics[width=0.45\textwidth]{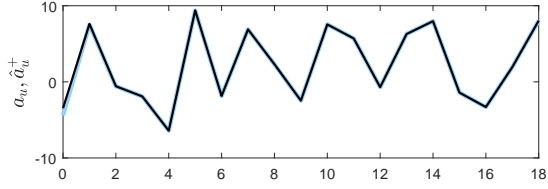}
		\caption{Estimated actuator attacks $\hat{a}_{u}^{+}$ converges to $a_{u}$ when $a_{u},a_{y3}\sim\mathcal{U}(-10,10)$. Legend: $\hat{a}_{u}^{+}$ (blue), $a_{u}$ (black).}
		\label{fig:5e}
		\centering
	\end{figure}

\begin{figure}[t]\centering
		\includegraphics[width=0.45\textwidth]{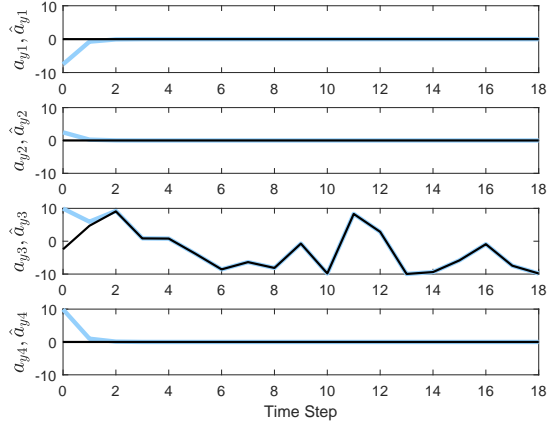}
		\caption{Estimated sensor attacks $\hat{a}_{y}$ converges to $a_{y}$ when $a_{u},a_{y3}\sim\mathcal{U}(-10,10)$. Legend: $\hat{a}_{y}$ (blue), $a_{y}$ (black).}
		\label{fig:6e}
		\centering
	\end{figure}

\begin{figure}[t]\centering
		\includegraphics[width=0.45\textwidth]{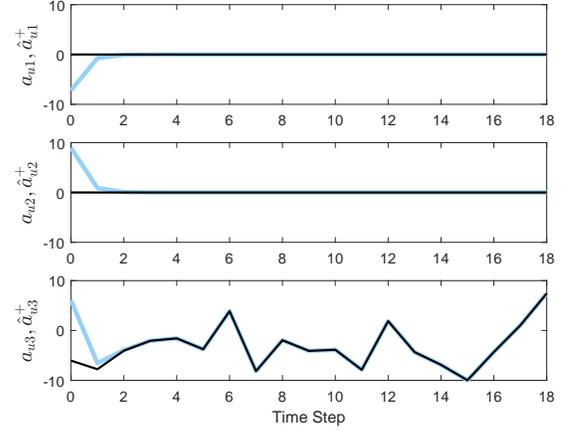}
		\caption{Estimated actuator attacks $\hat{a}_{u}^{+}$ converges to $a_{u}$ when $a_{u3},a_{y2}\sim\mathcal{U}(-10,10)$. Legend: $\hat{a}_{u}^{+}$ (blue), $a_{u}$ (black).}
		\label{fig:33}
		\centering
	\end{figure}

\begin{figure}[t]\centering
		\includegraphics[width=0.45\textwidth]{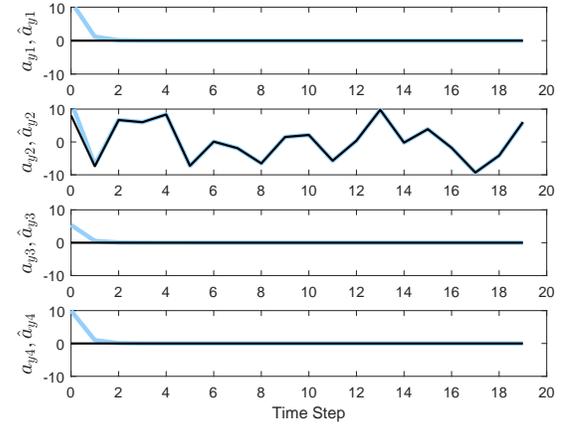}
		\caption{Estimated sensor attacks $\hat{a}_{y}$ converges to $a_{y}$ when $a_{u3},a_{y2}\sim\mathcal{U}(-10,10)$. $\hat{a}_{y}$ (blue), $a_{y}$ (black).}
		\label{fig:34}
		\centering
	\end{figure}

	Next, consider the partial multi-observer estimator given in Section \ref{partial}. In this case, the attack vector $a_{u}$ and $a_{y}$ can also be written as \eqref{attack_estimation_1} and \eqref{attack_estimation_2}, and the estimation error dynamics is given by some nonlinear difference equation characterized by the estimator structure in (\ref{54})-(\ref{56}). Let the estimation error dynamics be given by
	\begin{equation}\label{c4}
	e^{+}=f_{2}(e,x,a_{y},a_{u}),
	\end{equation}
	for some nonlinear function $f_{2}:\mathbb{R}^n \times \mathbb{R}^n \times \mathbb{R}^{n_{y}}\times\mathbb{R}^{n_{u}} \rightarrow \mathbb{R}^n$. In Theorem \ref{t1}, we have proved that $e$ converges to the origin exponentially. Hence, the attack estimate in \eqref{ru} and \eqref{ry} exponentially reconstructs the attack signals. Again, the sparsity pattern of $\hat{a}_{u}(k)$ and $\hat{a}_{y}(k)$ can be used to isolate actuator and sensor attacks using \eqref{wu} and \eqref{wy}.\\[1mm]
	\textbf{Example 3:}
	Consider system (\ref{e1}) and the complete multi-observer estimator in Example 1. Let $W_{u}=\left\lbrace 1\right\rbrace $, $W_{y}=\left\lbrace 3\right\rbrace $, $u\sim\mathcal{U}(-1,1)$, $a_{u},a_{y2}\sim\mathcal{U}(-10,10)$, and $(x_{1}(0),x_{2}(0))\sim\mathcal{N}(0,1^{2})$. We obtain $\hat{a}_{u}(k)$ and $\hat{a}_{y}(k)$ from (\ref{ru}) and (\ref{ry}). The reconstructed attack signals are depicted in Figures \ref{fig:5e}-\ref{fig:6e}. By checking the sparsity of these signals, actuator and sensor $3$ are isolated as attacked.
	\\[1mm]
	\textbf{Example 4:}
Here we consider system (\ref{e2}) and the partial multi-observer estimator in Example $2$. Let $W_{u}=\left\lbrace 3\right\rbrace $, $W_{y}=\left\lbrace 2\right\rbrace $, $(u_{1},u_{2},u_{3})\sim\mathcal{U}(-1,1)$, $a_{u3},a_{y2}\sim\mathcal{U}(-10,10)$, and $(x_{1}(0),x_{2}(0),x_{3}(0))\sim\mathcal{N}(0,1^{2})$. We obtain $\hat{a}_{u}(k)$ and $\hat{a}_{y}(k)$ from (\ref{ru}) and (\ref{ry}). The reconstructed attacks are shown in Figures \ref{fig:33}-\ref{fig:34}. In this case, using sparsity of the estimated attacks, actuator $3$ and sensor $2$ are correctly isolated.
	
\section{Control}\label{control}
In this section, we introduce a method to use the proposed multi-observer estimators to asymptotically stabilize the system dynamics.

\subsection{Sensor attacks only}
\begin{figure}[t]\centering
		\includegraphics[width=0.45\textwidth]{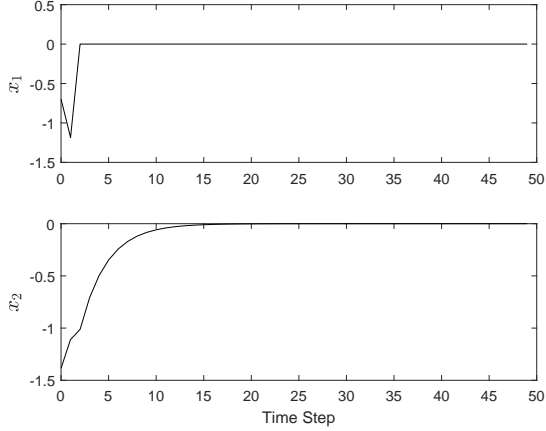}
		\caption{Controlled states when $a_{y2}\sim\mathcal{U}(-10,10)$. }
		\label{fig:10e}
		\centering
	\end{figure}

	We first consider the case when only sensors are attacked and actuators are attack-free. Then, the system is given by
	\begin{equation}\label{s4}
	\left\{\begin{split}
	x^{+}=&Ax+Bu,\\
	y=&Cx+a_{y}.
	\end{split}\right.
	\end{equation}
	Let $u=K\hat{x}$, where $\hat{x}$ is the estimate given by the complete multi-observer estimator in Section \ref{complete} or the partial multi-observer estimator in Section \ref{partial}, and $K$ is chosen such that $A+BK$ is Schur. Then, the closed-loop system is given by
	\begin{equation}\label{cc1}
	x^{+}=Ax+BK\hat{x},
	\end{equation}
or in terms of the estimation error as
	\begin{equation}
	\begin{split}\label{cc2}
	x^{+}=&Ax+B(K(\hat{x}-x+x)),\\
	=&(A+BK)x+BKe.
	\end{split}
	\end{equation}
	For the complete multi-observer estimator, let the estimation error dynamics be given by
	\begin{equation}
	\begin{split}\label{cc3}
	e^{+}=&f_{1}(e,x,a_{y}),\\
	\end{split}
	\end{equation}
	for some nonlinear function $f_{1}:\mathbb{R}^{n}\times\mathbb{R}^{n}\times\mathbb{R}^{n_{y}}\to\mathbb{R}^{n}$.
	For the partial multi-observer estimator, let the estimation error dynamics be given by
	\begin{equation}
	\begin{split}\label{cc4}
	e^{+}=&f_{2}(e,x,a_{y}),\\
	\end{split}
	\end{equation}
	for some nonlinear function $f_{2}:\mathbb{R}^{n}\times\mathbb{R}^{n}\times\mathbb{R}^{n_{y}}\to\mathbb{R}^{n}$.
	Since $A+BK$ is Schur, the closed-loop dynamics (\ref{cc2}) is Input-to-State Stable (ISS) with respect to input $e(k)$ and some linear gain, see \cite{Jiang2001}. Moreover, in Theorems \ref{th1} and \ref{t1}, we have proved that \eqref{cc3} and \eqref{cc4} are exponentially stable uniformly in $x(k)$ and $a_{y}(k)$. The latter and ISS of the system dynamics imply that $\lim_{k \rightarrow \infty}x(k) = 0$ \cite{Jiang2001}.\\[1mm]
	\textbf{Example 5:}
	Consider the open-loop unstable system
	\begin{equation}
	\left\{\begin{split}\label{e3}
	x^{+}=&\left[ \begin{matrix}
	1.2&0.5\\
	0.2&0.7
	\end{matrix}\right] x+\left[ \begin{matrix}
	1&0\\
	0&1\\
	\end{matrix}\right] K\hat{x},\\
	y=&\left[ \begin{matrix}
	1&1&3&2\\3&1&2&1
	\end{matrix}\right]^{\top} x+a_{y}.
	\end{split}\right.
	\end{equation}
	It can be verified that a UIO of the form \eqref{o} exists for each $J_{s}\subset\left\lbrace 1,2,3,4\right\rbrace $ with $\card(J_{2})\geq 2$; then, $4-2q=2$ and $q=1$. We let $W_{y}=\left\lbrace 2\right\rbrace $ and $a_{y2}\sim\mathcal{U}(-10,10)$. We construct $\binom{4}{3}+\binom{4}{2}=10$ UIOs initialized at $\hat{x}(0)=\left[ 0,0\right] ^{\top}$ and let \[K=\left[\begin{matrix}
	-1.2&0.7\\-0.2&-0.7
	\end{matrix}\right].\] We use the complete multi-observer in Section \ref{complete} to estimate the state. The state of the closed-loop system is shown in Figure \ref{fig:10e}.
	\subsection{Sensor and actuator attacks}
\begin{figure}[t]\centering
		\includegraphics[width=0.45\textwidth]{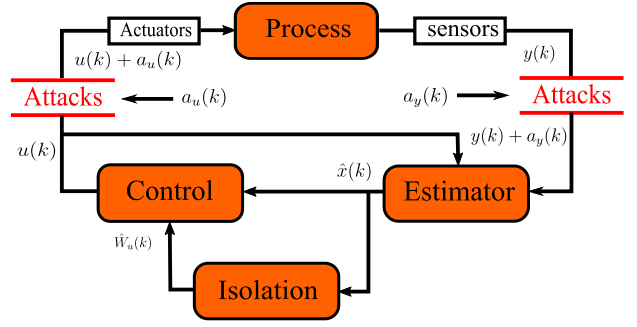}
		\caption{Estimation, isolation, and control diagram}
		\label{fig:11e}
		\centering
	\end{figure}

Here, we consider sensor and actuator attacks. We propose a simple yet effective technique to stabilize the system by switching off the isolated actuators, i.e., by removing the columns of $B$ that correspond to the isolated actuators, and closing the loop with a multi-observer based output dynamic feedback controller, see Figure \ref{fig:11e}. We introduce a switching signal $\rho(k)\subseteq\left\lbrace 1,\ldots n_{u}\right\rbrace $, containing the isolated attack-free actuators, i.e., $\rho(k) := \{1,\ldots,n_u\} \setminus \hat{W}_u(k)$. This $\rho(k)$ is used to denote actuators that are switched on. That is, $\rho(k)=J$ if the subset $J\subseteq\left\lbrace 1,\ldots,n_{u}\right\rbrace $ of actuators are switched on and the remaining actuators are switched off at time $k$. Again, let $B$ be partitioned as $B=\left[ b_{1},\hdots,b_{i},\hdots,b_{n_{u}}\right]$. After switching off the subset $\left\lbrace 1,\ldots,n_{u}\right\rbrace \setminus\rho(k)$ of actuators, system (\ref{s1}) is written as follows
\begin{equation}\label{s5}
	\left\{ \begin{split}
	x^{+}=&Ax+b_{\rho(k)}(u^{\rho(k)}+a_{u}^{\rho(k)}),\\
	y=&Cx+a_{y},
	\end{split}\right.
	\end{equation}
	 where $b_{\rho(k)}$ is the matrix whose columns are $b_{i}, i\in\rho(k)$, vectors $u^{\rho(k)}$ and $a_{u}^{\rho(k)}$ are the inputs and attacks corresponding to the switched-on actuators, respectively. We first consider the case when the complete multi-observer estimator in Section \ref{complete} exists, i.e., $\hat{x}$ is generated by (\ref{es1})-(\ref{es3}). We estimate $\hat{a}_{u}(k)$ using (\ref{ru}) and obtain $\hat{W}_{u}(k)$ from (\ref{wu}). Then, we switch off the set $\hat{W}_{u}$ of actuators by letting $\rho(k)=\bar{J}(k)=\left\lbrace 1,\ldots,n_{u}\right\rbrace \setminus\hat{W}_{u}(k)$. Since $a_{i}(k)=0, i\in\bar{J}(k)$, system \eqref{s5} has the following form:
	\begin{equation}\label{d1}
	x^{+}=Ax+b_{\bar{J}(k)}u^{\bar{J}(k)}
	\end{equation}
	where $u^{\bar{J}(k)}\in\mathbb{R}^{\card(\bar{J}(k))}$ is the set of isolated attack-free inputs. Let $0<q^{\star}<n_{u}$ be the largest integer such that $(A,b_{J})$ is stabilizable for each set $J\subset\left\lbrace 1,\ldots,n_{u}\right\rbrace $ with $\card(J)\geq n_{u}-q^{\star}$ where $b_J$ denotes a matrix whose columns are $b_i$ for $i \in J$. We assume that at most $q^{\star}$ actuators are attacked. It follows that $n_{u}-q^{\star}\leq\card(\bar{J}(k))\leq n_{u}$. We assume the following.
	\begin{assumption}\label{as3}
		For any subset $J$ with cardinality $\card(J)=n_{u}-q^{\star}$, there exists a linear switching state feedback controller $u^{\bar{J}(k)}=K_{\bar{J}(k)} x$ such that the closed-loop dynamics:
		\begin{equation}\label{c2}
		x^{+}=(A+b_{\bar{J}(k)}K_{\bar{J}(k)})x+b_{\bar{J}(k)}K_{\bar{J}(k)}e,
		\end{equation}
		is ISS with input $e$ for $b_{\bar{J}(k)}$ arbitrarily switching among all $b_{J'}$ with $ J\subset J'\subset\left\lbrace 1,\ldots,n_{u}\right\rbrace$ and $n_{u}-q^{\star}\leq\card(J')\leq n_{u}$.
	\end{assumption}
\begin{remark}
We do not give a method for designing the linear switching state feedback controller $u^{\bar{J}(k)}=K_{\bar{J}(k)} x$. Standard results for designing switching controllers, for instance results in \emph{\cite{Daafouz2002}} and references therein, can be used to design controllers satisfying Assumption \ref{as3}.
	\end{remark}
	By switching off the set $\hat{W}_{u}(k)$ of actuators at time $k$, using the controller designed for the set $\bar{J}(k)$, and letting $u^{\bar{J}(k)}=K_{\bar{J}(k)}\hat{x}$, the closed-loop system can be written as \eqref{c2} with estimation error $e=\hat{x}-x$ generated by some nonlinear difference equation (\ref{he}).
Because in Theorem \ref{th1}, we have proved that $e(k)$ converges to zero exponentially uniformly in $x(k)$, $a_{y}(k)$ and $a_{u}(k)$, the error $e(k)$ in (\ref{c2}) is a vanishing perturbation. Hence, under Assumption \ref{as3}, it follows that $\lim_{k \rightarrow \infty}x(k) = 0$.
	
\begin{figure}[t]\centering
		\includegraphics[width=0.45\textwidth]{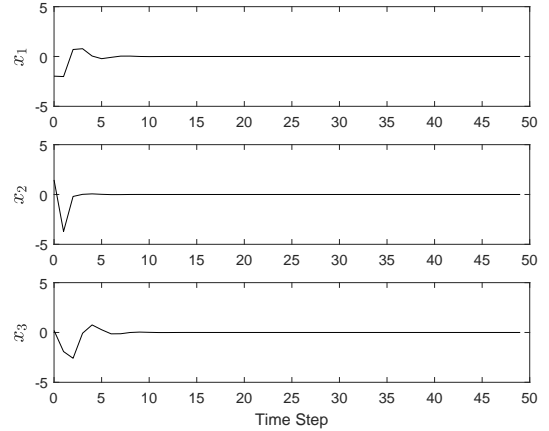}
		\caption{State trjectories when $a_{u3},a_{y2}\sim\mathcal{U}(-10,10)$.}
		\label{fig:12e}
		\centering
	\end{figure}

Next, assume that a complete multi-observer estimator does not exist but a partial multi-observer estimator exists (Section \ref{partial}), i.e., $\hat{x}$ is generated from \eqref{54}-\eqref{56} and $q_{1}\leq q^{\star}$. We assume that at most $q_{1}$ actuators are attacked. We construct $\hat{x}(k)$ from (\ref{54})-(\ref{56}), estimate $\hat{a}_{u}(k)$ using (\ref{ru}), and obtain $\hat{W}_{u}(k)$ from (\ref{wu}). After switching off the set $\hat{W}_{u}(k)$ of actuators, the system has the form (\ref{d1}) with $n_{u}-q_{1}\leq\card(\bar{J}(k))\leq n_{u}$. We assume the following.
	\begin{assumption}\label{as4}
		For any subset $J$ with cardinality $\card(J) = n_{u}-q_{1}$, there exists a linear switching state feedback controller $u^{\bar{J}(k)}=K_{\bar{J}(k)} x$ such that the closed-loop dynamics \eqref{c2} is ISS with respect to $e$ for $b_{\bar{J}(k)}$ arbitrarily switching among all $b_{J'}$ with $ J\subset J'\subset\left\lbrace 1,\ldots,n_{u}\right\rbrace$ and $n_{u}-q_{1} \leq\card(J')\leq n_{u}$.
	\end{assumption}
	Using the controller designed for the set $\bar{J}(k)$, and letting $u^{\bar{J}(k)}=K_{\bar{J}(k)}\hat{x}$, the closed-loop dynamics can be written in the form (\ref{c2}). Then, in this case, $e(k)$ is generated by some nonlinear difference equation of the form \eqref{c4}. Under Assumption \ref{as4}, the closed-loop dynamics (\ref{c2}) is ISS with input $e(k)$, see \cite{Jiang2001}. Moreover, in Theorem \ref{t1}, we have proved that $e(k)$ converges to the origin exponentially uniformly in $x(k)$, $a_{u}(k)$ and $a_{y}(k)$. The latter and ISS of the system dynamics imply that $\lim_{k \rightarrow \infty}x(k) = 0$ \cite{Jiang2001}.\\[1mm]
	\textbf{Example 6:}\label{ex7}
	Consider the following system:
	\begin{equation}
	\left\{\begin{split}
	x^{+}=&\left[ \begin{matrix}\label{e7}
	0.5&0&0.1\\
	0.2&1.7&0\\
	1&0&0.3
	\end{matrix}\right] x+\left[ \begin{matrix}
	0.5&0&1\\
	1&1&1\\
	0&0&1
	\end{matrix}\right] (u+a_{u}),\\
	y=&\left[ \begin{matrix}
	1&2&0\\0&1&1\\0&1&2\\1&1&1
	\end{matrix}\right] x+a_{y}.
	\end{split}\right.
	\end{equation}
	 Since $(A,b_{i})$ is stabilizable for $i\in\left\lbrace 1,2,3\right\rbrace $, we have $q^{\star}=2$. It can be verified that there does not exist a complete UIO for any $S_{s}\subset\left\lbrace 1,2,3,4\right\rbrace $ with $\card(S_{s})=2$, but partial UIOs exists for each pair $(J_{u},J_{s})$ with $\card(J_{u})\leq 2$ and $\card(J_{s})\geq 2$; then, we have $q_{1}=q_{2}=1$ and $q_{1}<q^{\star}$. We let $W_{u}=\left\lbrace 3\right\rbrace $, $W_{y}=\left\lbrace 2\right\rbrace $, and $a_{u3},a_{y2}\sim\mathcal{U}(-10,10)$. We construct $\binom{3}{1}\times\binom{4}{3}+\binom{3}{2}\times\binom{4}{2}=30$ UIOs and use the design method given in \cite{Daafouz2002} to build controllers for actuators $\left\lbrace 1,2\right\rbrace $, $\left\lbrace 1,3\right\rbrace $, $\left\lbrace 2,3\right\rbrace $, $\left\lbrace 1,2,3\right\rbrace $. Then, we use the partial multi-observer approach in Section \ref{partial} to estimate the state, reconstruct the attack signals and control the system. The state of the system is shown in Figure \ref{fig:12e}.
	\section{Conclusion}
	
	We have addressed the problem of state estimation, attack isolation, and control for discrete-time linear time-invariant (LTI) systems under (potentially unbounded) actuator and sensor false data injection attacks. Using a bank of Unknown Input Observers (UIOs), we have proposed an estimator that reconstructs the system states and the attack signals. We use these estimates to isolate attacks and control the system. We propose an effective technique to stabilize the system by switching off the isolated actuators. Simulation results are provided to illustrate our results.
	
	
	
	
	\bibliographystyle{ieeetr}
	\bibliography{Observer}
	%
	%
	%
\end{document}